\documentclass{aa}

\usepackage{epsf}

\usepackage{times}

\begin{document}

   \thesaurus{06
              (08.09.2 \object{1WGA J1958.2+3232};
           08.14.2;
               08.02.1;
               13.25.5;
               03.09.4)}
   \title{ Periodic modulation of the optical counterpart of the X-ray pulsator 1WGA J1958.2+3232.
   A new intermediate polar.\thanks{Based on observations collected at Catania and Asiago Astrophysical
Observatories.} }

   \author{ M. Uslenghi\inst{1} \and
        P. Bergamini\inst{1} \and
        S. Catalano\inst{2} \and
        L. Tommasi\inst{1,3} \and
        A. Treves\inst{3}
          }

   \offprints{ M. Uslenghi (uslenghi@ifctr.mi.cnr.it)}

   \institute{ Istituto di Fisica Cosmica "G.Occhialini",CNR,
              Via Bassini 15, I-20133 Milano, Italy.
         \and
Osservatorio Astrofisico di Catania, Via S.Sofia, 78, I-95125
Catania, Italy.
       \and
Universit\`a degli Studi dell'Insubria, Polo di Como,
Dipartimento di Scienze Chimiche, Fisiche e Matematiche, Via
Lucini 3, 22100, Como, Italy.
             }

   \date{Received September 00, 0000; accepted March 00, 0000}
   \titlerunning{Optical modulation of \object{1WGA J1958.2+3232}}
   \authorrunning{M. Uslenghi et al.}
   \maketitle

   \begin{abstract}

Time-resolved observations of a 6' x 6' field, containing the
position error boxes of the X-ray source \object{1WGA
J1958.2+3232}, were performed in June 1999 with the 91 cm Catania
telescope, equipped with a Photon Counting Intensified CCD. The
star recently proposed as the optical counterpart of the
\object{1WGA J1958.2+3232}, exhibited a strong optical modulation
with a period compatible with that seen in X-ray ($\sim$~12~min).
The optical modulation was detected again in September and October
1999. These results confirm the identification of the optical
counterpart and support the classification of \object{1WGA
J1958.2+3232} as a cataclysmic variable, possibly an
\textit{Intermediate Polar}. Modulation at period twice as large
was also found in one observation run, suggesting that the true
spin period of the White Dwarf could be 24 min rather than 12
min, thus \object{1WGA J1958.2+3232} appears to be, among the IPs,
the slowest rotator which exhibit double peaked spin profile.

   \end{abstract}

%

\section{Introduction}

\object{1WGA J1958.2+3232} is one of the objects found by Israel
et al. (\cite{Isr98}) in a systematic search for pulsators in the
catalogue of ROSAT X-ray sources compiled by White, Giommi \&
Angelini (\cite{WGA94}). The source appeared with a mean flux (in
0.1-2.4 keV) of $\sim$ 10$^{-12}$ erg cm$^{-2}$ s$^{-1}$,
modulated with a period 721 $\pm$ 14 s and pulsed fraction of
about 80\%. A subsequent observation performed by ASCA confirmed
both the flux level and the strong periodic modulation at 734
$\pm$ 1 s (Israel et al. \cite{Isr99}). The energy spectrum, as
measured by ROSAT, is consistent with a power law model with a
photon index $\Gamma$=0.8$^{+1.2}_{-0.6}$ and a column density
N$_{H}$=(6$^{+24}_{-5}$)$\times$10$^{20}$cm$^{-2}$.

On the basis of optical photometry and spectroscopy, Israel et
al. (\cite{Isr99}) proposed a m$_V$=15.7 star as the counterpart
of  \object{1WGA J1958.2+3232}. The spectrum of this star,
characterized by strong emission lines, was classified as B0Ve
type, suggesting that \object{1WGA J1958.2+3232} could be a
Be/X-Ray binary with an accreting neutron star. The assumed
distance of 800 pc (resulting from interstellar absorption
spectral features) yielded an X-ray luminosity of
8$\times$10$^{31}$ erg s$^{-1}$ in 0.1-2.4 keV.

Recently Negueruela et al. (\cite{Neg99}) reported new
spectroscopic measurements, with a higher signal to noise ratio,
of the Israel et al. (\cite{Isr99}) candidate. They found that
optical properties of the source are incompatible with a Be/X-Ray
binary, because photospheric features are absent, whereas some
characteristics of the spectrum, such as the strong emission in
all Balmer lines, are typical of a cataclysmic variable. The
shape of the lines, strongly asymmetric and double peaked,
suggests the presence of an accretion disc.


We included \object{1WGA J1958.2+3232} in our target list during
a campaign for scientific performance evaluation of a photon
counting detector that we carried out in June 1999. We report
here on the results of this campaign and of subsequent
observations.


\section{Observations}
The observations were carried out with the 91 cm telescope of the
Astrophysical Observatory of Catania (OACT)~-~Serra la Nave and
with the 182 cm telescope of the Astronomical Observatory of
Padua (OAPD)~-~Asiago Cima Ekar (see the observations log in
Table~\ref{TabOss}). Two kinds of detectors were used: a photon
counting camera based on an Intensified CCD\footnote{The ICCD
detector has been developed at the Istituto di Fisica Cosmica
"G.Occhialini", in collaboration with Dipartimento di Ingegneria
Elettronica of Padua University.} (Bergamini et al.
\cite{Ber97,Ber98,Ber00}) and a standard photomultiplier
photometer.

The Photon Counting ICCD (PC-ICCD) employs a high-gain
Micro-Channel Plate (MCP) image intensifier as input device. The
intensifier converts virtually each photon, interacting
photoelectrically with the photocathode, into a luminous spot on
a phosphor screen that preserves the \textit{(x,y)} location of
the event. The photon event is reimaged onto a fast framing CCD
camera by means of a fiber optic taper with a 3.6:1
demagnification ratio. The camera detects the location of each
event and a dedicated digital electronic system establishes the
coordinates of the centroid of the events with subpixel accuracy.
The detector can operate in image integration mode, where the
collected photons are accumulated in a memory array, or in photon
list mode where the photon events are recorded as a list of time
tagged event coordinates with temporal resolution as low as 4.5
ms.

Thus, in photon list mode, the PC-ICCD is able to perform high
speed space resolved photometry, allowing to investigate time
variability down to few ms, simultaneously for all the stars in
the field of view.

The PC-ICCD was developed for space UV astronomy and uses a RbTe
photocathode: even though the photocathode shows its peak
sensitivity in the vacuum UV range, it yields a residual quantum
efficiency ($\leq$ 0.1 \%) for ground based observations
(Bergamini et al. \cite{Ber00}). In order to preserve this
residual efficiency, no photometric filter has been used with the
ICCD; the boundaries of instrumental band (3200-5500 \AA) were set
respectively by the atmosphere UV cut-off and by the detector
photocathode sensitivity.

For measurements in photometric bands, we used a photon-counting
cooled photometer of OACT, equipped with an EMI 9893QA/350
photomultiplier.

\begin{table*}
      \caption[]{Observations Log.}
\begin{tabular}{ l  c  c  c  l  l  c  c  }

\hline \hline
Date &Detector   &Site &Telescope &UT start &Duration  &Time resolution &Filter$^{\mathrm{a}}$ \\

\hline
June 9, 1999 &ICCD &OACT &91 cm  &01.27.10 &2686 s &16.8 ms &WL \\
June 13, 1999 &ICCD &OACT &91 cm  &01.27.07 &3600 s &16.8 ms &WL \\
June 15, 1999 &ICCD &OACT &91 cm  &01.18.15 &3768 s &16.8 ms &WL \\
June 17, 1999 &ICCD &OACT &91 cm  &00.58.42 &5004 s &16.8 ms &WL \\
June 18, 1999 &ICCD &OACT &91 cm  &00.14.20 &7565 s &16.8 ms &WL \\
September 7, 1999 &EMI 9893QA/350 &OACT &91 cm  &21.03.02 &7000 s &5 s &U \\
September 14, 1999 &EMI 9893QA/350 &OACT &91 cm  &20.02.38 &9000 s &5 s &U \\
October 1, 1999 &ICCD &OAPD &182 cm  &21.14.01 &9646 s &4.512 ms &WL \\
\hline

\end{tabular}
\begin{list}{}{}
\small {\item[$^{\mathrm{a}}$] WL = White Light }
\end{list}

         \label{TabOss}
      \vspace{0.5cm}

   \end{table*}

\section{Data analysis and results}
\subsection{Source identification and short term variability
} \label{hairymath}

On several nights in June 1999, \object{1WGA J1958.2+3232} was
observed in photon list mode with a time resolution of 16.8 ms
for a total of  6.3 h distributed in five integrations
(Table~\ref{TabOss}). The 27.6 $\times$ 27.6 mm$^2$ sensitive
area of the PC-ICCD covered a 6.4' $\times$ 6.4' field of view
with an effective resolution of 0.35 arcsec (FWHM).

\begin{figure}[htbp]
  \begin{center}
    \leavevmode
     \epsfxsize=8.5cm  \epsfbox{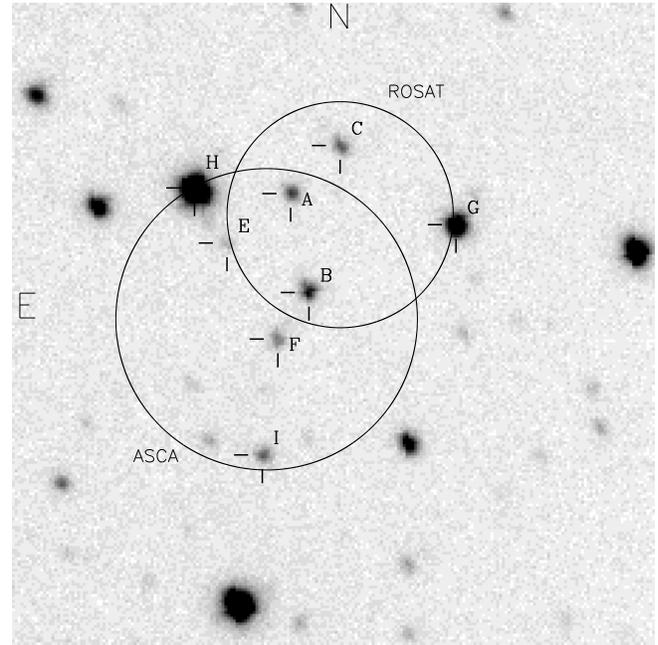}
    \caption{The field of \object{1WGA J1958.2+3232} accumulated from
    the five observing runs in June 1999. Both ROSAT and ASCA position
    error circles (90\% confidence) are shown. The eight stars labeled
    were searched for periodicity. }
    \label{fig1}
    \end{center}
\end{figure}

A set of IDL routines were used to process the data stream of
each observation and to perform data reduction adopting the
following procedure. The observations were segmented into 60 s
long data subsets. From each subset a 512 $\times$ 512 image
array (pixel size 0.75$\arcsec$) was accumulated and flat--field
corrected. In order to add consistently the field images acquired
in the five separate observation runs and to compensate for small
tracking inaccuracies within the single run, we selected a bright
isolated source and used its centroid to coalign the 60 s
images.  The latter procedure, which improved the overall Point
Spread Function (FWHM) from 4$\arcsec$ to 3$\arcsec$, was also
applied to the light curve extraction process described below. A
2.8$\arcmin\times$2.8$\arcmin$ section of the field accumulated
from the 6.3~h observation is shown in Fig.~\ref{fig1}: the
circles represent the position uncertainty (90\% confidence) of
ROSAT (30$\arcsec$) and ASCA (40$\arcsec$) X-ray detections.

From the sources found inside, or close to, the two position
error circles we selected eight objects detected well above the
sky background to be searched for variability. Note that the
contribution of the detector noise to the background is virtually
nil, being as low as 10$^{-5}$ counts~s$^{-1}$~pix$^{-1}$ (sky
background was $\sim$~10$^{-2}$ counts~s$^{-1}$~pix$^{-1}$). From
each object labeled in Fig.~\ref{fig1} (the candidates of Israel
et al. \cite{Isr99} are reported with the same name, except star
D which is too faint in our data) we extracted the source and
background light curve binned at 1 s resolution. Great care was
paid in the selection of the aperture radii used in the process,
the field of \object{1WGA J1958.2+3232} being located in a
crowded region near the galactic equator. The optimal source
aperture radius was found at 2.25$\arcsec$ whereas for the
background a concentric ring of inner and outer radii
respectively measuring 9$\arcsec$ and 11.25$\arcsec$ was used.
With the same procedure the light curves of three bright stars
found well outside both ROSAT and ASCA error box were generated
and, since a preliminary time series analysis excluded any
intrinsic periodicity, they were used as comparison to remove any
atmospheric effect.

\begin{figure}[htbp]
  \begin{center}
    \leavevmode
     \epsfxsize=8cm  \epsfbox{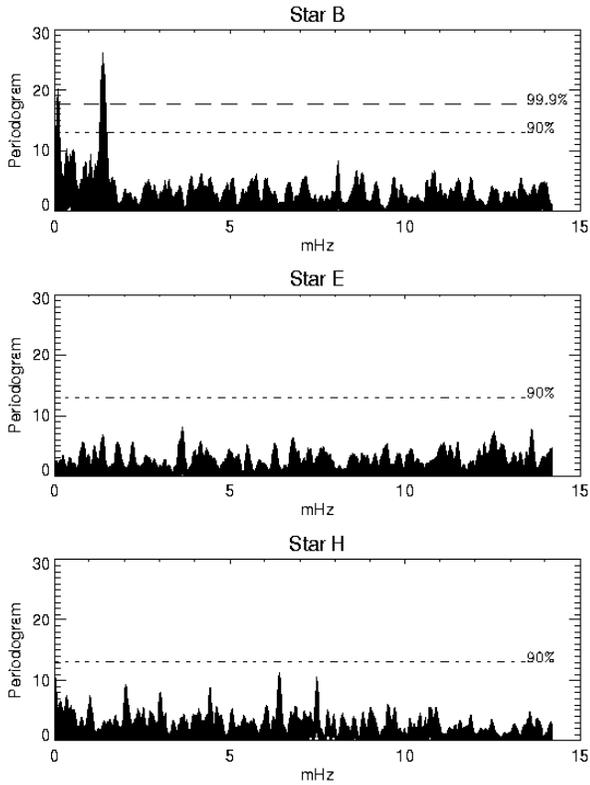}
    \caption{ Lomb-Scargle periodograms of the light curves
    of three selected stars. The periodogram was computed over
    the whole time span of the observations. The dashed and dotted
    lines respectively show the $90 \%$ and $99.9 \%$ confidence level
    in the hypothesis of random noise.}
    \label{fig2}
    \end{center}
\end{figure}

To investigate the time variability, for each object we computed
a Lomb-Scargle periodogram (Scargle \cite{Sca82}) on the combined
data set. Some typical periodograms are shown in Fig.~\ref{fig2}.
Only star B shows a significant periodic signal. The periodogram
shows two peaks, the most apparent corresponding to about 721 s
period, consistent with the X-ray period. Star B is the source
previously proposed as optical counterpart of \object{1WGA
J1958.2+3232} by Israel et al. (\cite{Isr99}).

In order to look for longer periodicity, we planned new
observations with the PC-ICCD at Asiago Observatory at the
beginning of October but, due to poor weather conditions, we were
able to collect only about 2$^{h}$40$^{m}$ of data. The detector
was operated in the highest time resolution mode (4.512 ms),
corresponding to the widest dynamic range, but with a reduced
field of view (a quarter of the sensitive area of the detector,
resulting, with the Asiago plate scale, in 5.8' $\times$ 1.45').
This choice was determined by the necessity to preserve, under
higher count rates resulting from the better efficiency of the
telescope, the linearity of the detector also for the brightest
nearby stars, employed as comparison in data reduction of June
observations. Thus, we were able to use the same comparison stars
also for the October 1 run.

Observations in standard optical photometric bands have been
carried out with the photoelectric photometer, again at the 91 cm
telescope of the Catania Astrophysical Observatory. The
photometer was operated in two different mode: i) in the standard
mode to determine the U B V magnitudes and colours in the Johnson
system, ii) in continuous acquisition at fixed filter,
the  U filter in our case. A diaphragm of $\phi$~=~15 arcsec was adopted.\\
The V magnitude and colour indices of \object{1WGA J1958.2+3232}
were determined calibrating on nearby field stars of known
magnitudes and colours (e.g. \object{HD 188992} V=8.29,
B-V=-0.07, U-B=-0.38, and \object{HD 189596} V= 7.54, B-V=-0.11,
U-B=-0.44). Average seasonal atmospheric extinction  and
instrumental coefficients have been adopted to transform
instrumental magnitudes to the Johnson system. The magnitudes and
colours we obtain for \object{1WGA J1958.2+3232} are V= 15.713
$\pm$ 0.073, B-V  = 0.231 $\pm$ 0.067, U-B = -0.784 $\pm$ 0.096.

Continuous monitoring observations on September 7 and 14, with
the U filter and integration time of 5 sec for each measurement,
have been done with the aim of improving the measured accuracy of
the 12 m period. Uninterrupted sequences of 7000 second and of
9000 sec were obtained in September 7 and 14 respectively. The
sky background was subtracted by linear interpolation of the
values measured on ten samplings obtained at the beginning and at
the end of each run.  Correction for the atmospheric extinction
have been made by adopting the average seasonal coefficient for
the U filter.

No measurements of the sky background or of comparison stars have
been made during the observing runs, to avoid interruptions of
the data string and introduction of spurious frequencies in the
power spectrum. This prevents us from using these data to study
the long term variability.

\begin{figure}[htbp]
  \begin{center}
    \leavevmode
     \epsfxsize=8.5cm  \epsfbox{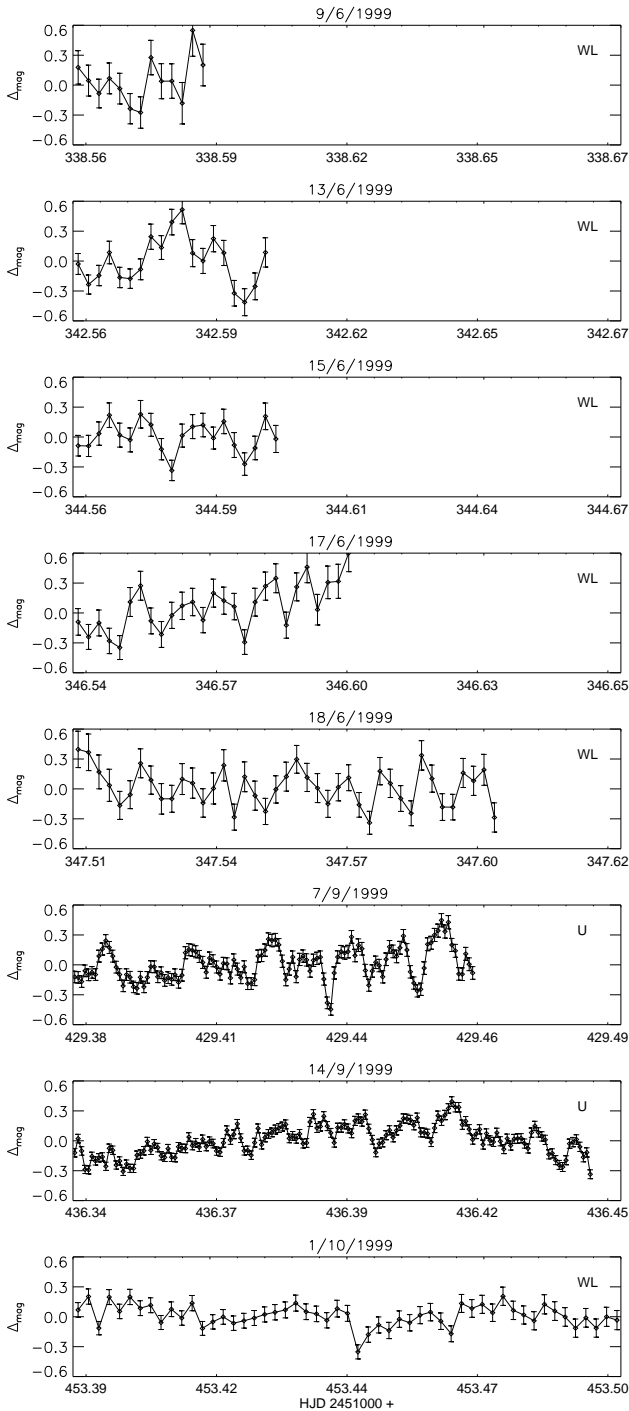}
    \caption{ \object{1WGA J1958.2+3232} light curves.
    Binning time: 180 s for 'white light photometry' and 60 s for
    U photometry.}
    \label{fig3}
    \end{center}
\end{figure}

The light curves of the source in the single nights, reduced as
previously explained, are shown in Fig.~\ref{fig3}.

\begin{figure}[htbp]
  \begin{center}
    \leavevmode
     \epsfxsize=8.5cm  \epsfbox{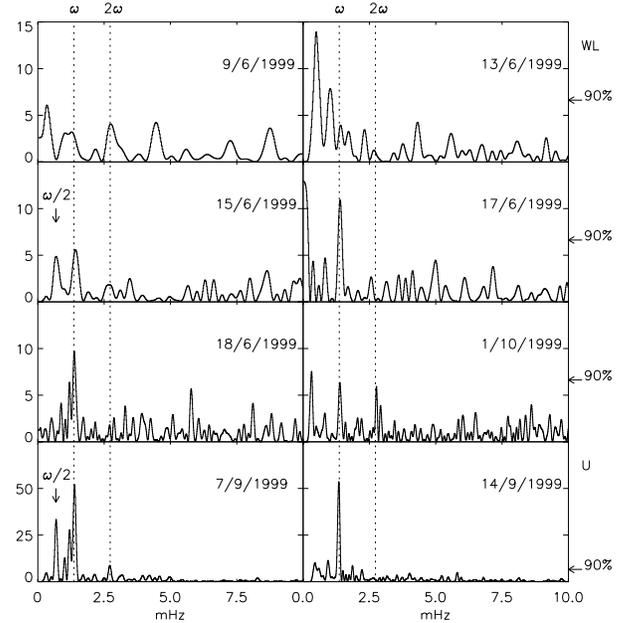}
    \caption{ Lomb-Scargle periodograms of star B for the
    eight observing sessions. The horizontal arrows show the
    $90\%$ confidence level while the vertical lines mark the
    frequency $1/733.24 s$ and its first harmonic. In 15 June and
    7 September periodograms the peaks corresponding to 24~min
    period are also indicated by vertical arrows.}
    \label{fig4}
    \end{center}
\end{figure}

A Lomb-Scargle periodogram was applied on the time series from
separate nights, after removing low frequency trends in U band
observations (carried out without comparison stars) by a
polynomial fit on each data set. The periodograms are shown in
Fig.~\ref{fig4}: the best period and its first harmonic are
marked by vertical dotted lines. In all but one observations a
peak was clearly visible near the period of the X-ray modulation.
The amplitude of the peak varies from night to night, but in the
longest runs it is statistically significant. On the other hand,
in the periodogram of June 13, there is only a hint of a peak.
The different run length doesn't justify the difference in the
observability of the modulation: the peak of the longest run
(October 1) is well below the 90\% confidence level. In four
observations a weaker peak is also present near the first
harmonic of the X period, suggesting that pulse shape is time
dependent and not simply sinusoidal. We notice also the presence
of a peak corresponding to a period about twice the main period
that can be seen in the periodograms of June 15 and September 7.
In September 7, the confidence level for this frequency is
greater than 99.999 \%.

\begin{figure}[htbp]
  \begin{center}
    \leavevmode
     \epsfxsize=8.5cm  \epsfbox{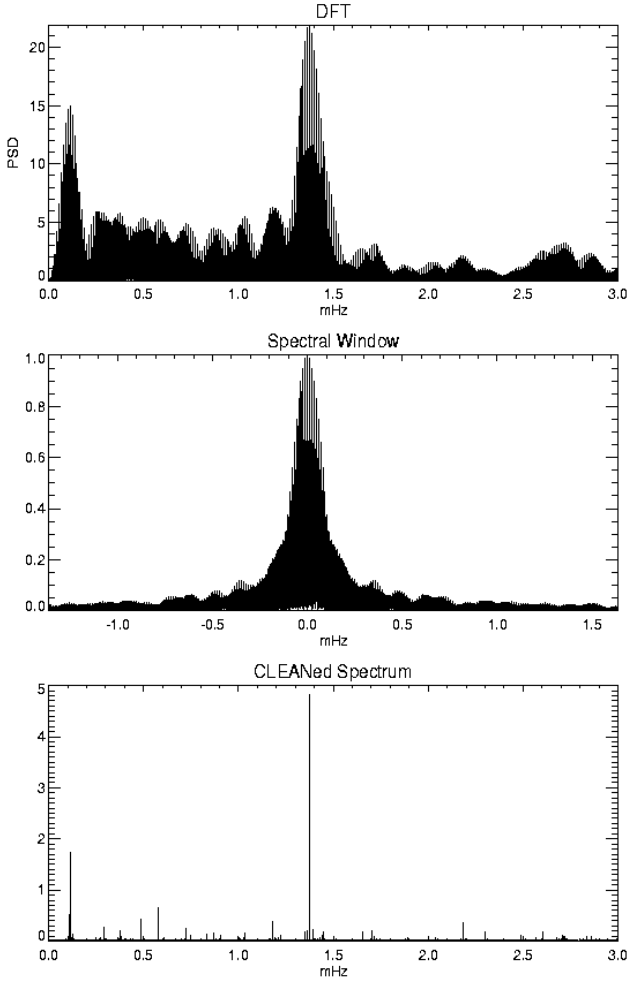}
    \caption{Discrete Fourier Transform of the light curve of star B.
(a) Power Spectrum, (b) Spectral Window, (c) CLEAN-ed spectrum
(see text).}
    \label{fig5}
    \end{center}
\end{figure}

In order to improve the resolution in the main period measure, a
Discrete Fourier Transform (DFT) was computed on the overall
observation set (Deeming \cite{Dee75}), excluding the observation
of October 1, because the modulation was very low. The spectrum
is shown in Fig.~\ref{fig5}, together with the spectral window.
The clean spectrum, produced by the deconvolution of the raw
spectrum with the spectral window by means of the CLEAN algorithm
(Roberts et al. \cite{Rob87}), is also included in the lower
panel of Fig.~\ref{fig5}. The resulting main period is 727.06
$\pm$ 0.02 s, but, due to the complex spectral window, dominated
by the 1/d aliasing, there is an ambiguity as to which of the
alias peak is the true period. A MonteCarlo simulation showed
that 1/d alias 721 and 733 s are acceptable (the resulting
probability for the 727 s peak to be the true period is about 70
\%, whereas there is about 30 \% probability for either 721 s or
733 s peak). In particular, 733.24 $\pm$ 0.02 is compatible with
the ASCA measure (734 $\pm$ 1 s), thus we consider this period as
the most likely. In the CLEANed spectrum it is also apparent the
presence of some weaker peaks,near 833~s and at some others lower
frequency.

\begin{figure}[htbp]
  \begin{center}
    \leavevmode
     \epsfxsize=8.5cm  \epsfbox{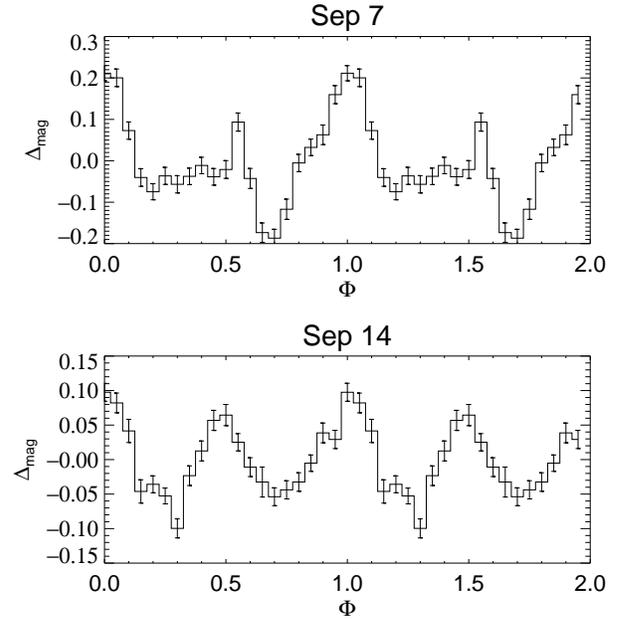}
    \caption{ Light curves from the two U photometric observations,
    folded with the 1466.48 s period, with 20 bins, relative to
    the mean level in each night. Initial phase is arbitrarily
    selected for September 7 data, and is maintained for September 14.}
    \label{fig6}
    \end{center}
\end{figure}

Instead, there is no significant peak near the half frequency of
the main modulation. The FFT computed on the longest data set in
which this modulation is apparent yields a best period
determination of 1448$\pm$158 s. The DFT computed on both the
data set showing this modulation doesn't improve the accuracy. In
fact, the June 15 observation is far shorter, with lower
statistics and the large gap between the two observations results
in a quite complex spectral window. In order to better estimate
the period, the epoch folding technique (Leahy et al.
\cite{Lea83}) was applied to the September 7 data, after removing
the 733.24 modulation by subtracting a Fourier fit (in the epoch
folding diagrams peaks are found at all the sub-harmonics of the
signal frequencies). The resulting period was 1467 $\pm$ 25 s (the
uncertainty has been derived from simulations), consistent with 2
$\times$ 733 s. After subtracting the long term variation, the
light curves of September 7 and 14 were folded at twice the
733.24 s period (Fig.~\ref{fig6}). A visual inspection of the
folded light curves confirms the presence of an even-odd effect,
more evident in the September 7 pulse shape.

\subsection{ The long term variability
} \label{hairymath}

Low frequency, high amplitude variations are clearly apparent in
the longest differential photometry runs and from night to night.
For example, the mean flux on October 1 is 50\% lower than the
mean flux averaged on the five nights of June. Moreover, on June
17, the source intensity increases by about 50\% during
$\sim$~1$^{h}$20$^{m}$. In both observations carried out with the
photometer, without comparison stars, the data show long term
trends not easily attributable to instrumental or transparency
changes. The data of September 7 show a slightly increasing trend
from the beginning to the end of the observations, even before the
correction for the atmospheric extinction is applied. After the
atmospheric
extinction correction, the average increase is about  15 \%.\\
The data of September 14 show a steady increase in the first half
of the observations, up to about the 40 \%, and then a decrease
to about the initial level. Although the changes in  September 7
night could result from an improvement of the sky transparency,
the trend observed on September 14 is more difficult to explain
in the same way. In both cases the observations were made after
the object meridian passage, therefore, the extinction effect
should increase with time, contrary to the observed trends. The
sky was very clear on both nights, so that an increase of the
transparency of about the 40 \%,  as shown by the data of the
second night, is very unlikely, also because the subsequent
observations made during the rest of the night do not support
this possibility. Changes on the photometer gain are excluded by
the stability of the photon counting system over several years.
Also changes in the sky background can be excluded because we
were in dark period, and the observing direction is free of any
close city light. The telescope was manually guided all the time
through an intensified TV camera using an off-set nearby star.
Normally the correction of the telescope wiggle produces random
noise, or loss of signal, but no long term trends. We are,
therefore, inclined to attribute the observed long term variation
to real variations of intensity of \object{1WGA J1958.2+3232}.

A peak is apparent in the DFT at low frequency (the second
highest peak in the CLEANed spectrum showed in Fig.~\ref{fig5}),
corresponding to a period of 2.45$^{h}$ (with alias at 2.73$^{h}$
and 2.23$^{h}$) with modulation of about 7.5\%. However, due to
the poor sampling, we are not able to reliably ascribe it to an
orbital period. In fact the period corresponding to the peak is
only slightly shorter than the two longest runs.

\section{Discussion}

Our observed B-V and U-B color indices are not compatible with
colors of any unreddened {\it normal} stellar object. If it was a
reddened object, adopting the Cygnus reddening law $E(U-B)=0.9
\cdot E(B-V)$, \object{1WGA J1958.2+3232} would come out to be a
hot star of spectral type about O5, with  E(B-V) $\sim$~0.6.
Assuming a standard reddening law, the visual interstellar
extinction $A_V=3.1 \cdot E(B-V) = 1.86$~mag would lead to a
distance of 1.5-2 Kpc, not compatible with the absence of
interstellar absorption feature in Na I D2 lines and the weak
interstellar feature observed in the Ca II K line (Negueruela et
al. \cite{Neg99}), which are in agreement with a far lower
distance, not above 500 pc.

In fact \object{1WGA J1958.2+3232} in the color-color diagram
lies on the Black-Body and white dwarf region. According to our
measurements (B-V = 0.231 $\pm$ 0.067 and U-B = - 0.784 $\pm$
0.096), it falls a little above the B-B line and the main
distribution of DA white dwarfs: it should be an object of about
T=10,000 K whose optical emission comes from the disc accretion
region.

Large modulation at the 733 s period (the same as measured in
X-ray) was found in almost every night, even if with a variable
amplitude. Weaker power peaks are present only in two data sets,
but further measures are needed to confirm their significance.
The detection of periodical optical modulation at the X-ray
period rules out any possibility that \object{1WGA J1958.2+3232}
could be a binary system hosting a Be, because the optical
emission should be dominated by the early type star, whereas the
modulated component comes from the compact object. A stable
optical period, usually in the range from some tens of seconds to
some tens of minutes, and X-ray modulation at or close to the
same period, is a signature of the Intermediate Polar class of
Cataclysmic Variables.

Generally, in the Intermediate Polar power spectra, the main
power is found at the spin period in X-ray (with the exception of
\object{RX1712-24}, Buckley et al. \cite{Buc97}), whereas the
optical light curves can be dominated by the spin (e.g.
\object{V709 Cas}, Norton et al. \cite{Nor99}, \object{BG CMi},
De Martino et al. \cite{DM95}, \object{UU Col}, Burwitz et al.
\cite{Bur96}), or by a sideband (e.g. \object{AO Psc}, Hellier et
al. \cite{Hel91}, \object{V1223 Sgr}, Jablonski \& Steiner
\cite{Jab87}, \object{RX J1712-24}, Buckley et al. \cite{Buc95}),
or by the orbital modulation, depending on the strength of the
magnetic field and the accretion geometry and rate. In
particular, differences in the accretion rate can produce
variation in the dominant frequency (e.g. \object{FO Aqr}, which
in the past was dominated by the spin period, De Martino et al.
\cite{DM94}, recently was found with the orbital modulation
dominating over all the other periods, De Martino et al.
\cite{DM99}).

In the X-Ray light curve measured by ROSAT, Israel et al.
(\cite{Isr98}) found modulation at the same period that we see in
optical and at its first harmonic, whereas they didn't find any
evidence of a period shorter then the main one. Thus, the
dominant period in X-rays and the optical should be the spin of
the white dwarf, rather than a sideband of a shorter period. The
dominance of the spin period and the non-detection of sidebands
neither in X-ray nor in optical light curves suggest that
\object{1WGA J1958.2+3232} is a disc accretor.

The amplitude of the modulation varies greatly from night to
night, but this is a rather common behavior for the Intermediate
Polars class (see, for example, Welsh \& Martell \cite{Wel96}).
Especially in the optical domain, IPs often show night to night
variations in the relative amplitudes of the power spectrum at
the characteristic frequencies. The presence of power in the DFT
at the first harmonic of the 12~min period, in some nights only,
suggests that the pulse shape could be variable.

The presence of another peak at half frequency suggests that the
true spin period could be 1466~s: then, the 12 min period would
be the first harmonic and the peak sometimes seen at 368$\pm$7 s
would correspond to the third harmonic. Among the Intermediate
Polars there are three systems (\object{DQ Her}, \object{YY Dra}
and \object{V405 Aur}, Allan et al. \cite{All96}) which have
optical and X-ray light curves dominated by the modulation at the
first harmonic of the spin period, whereas at the fundamental
frequency the modulation is very weak, or absent. However, all
these objects are characterized by a fast spinning white dwarf
(142 s, 529 s, 545 s respectively, Ritter \& Kolb \cite{Rit98}).
Commonly, the double peak pulse shape is interpreted as a
signature of two poles accretion, but Norton et al.
(\cite{Nor99}) point out that two poles accretors generally
produce single peaked profiles (e.g. in \object{EX Hya} there is
strong evidence of two poles accretion but the pulse profile is
single peaked, Hellier \cite{Hel95}). They outline a model
according to which, among the two poles accretors, only short
period IPs (with a period below 700 s) could exhibit a double
peaked pulse profile. For another system (\object{BG CMi}) there
has been some uncertainty whether the dominant period (913 s) is
the spin period or a sideband of its first harmonic (Patterson \&
Thomas \cite{Pat93}), but successive observations did not find
any evidence confirming the longer 1693 s period (Garlick et al.
\cite{Gar94}, De Martino et al. \cite{DM95}, Hellier
\cite{Hel97}).

If 24 min were the true period, \object{1WGA J1958.2+3232} would
be the slowest rotator which exhibits double peaked profile. In
the Norton model it should have a single peaked pulse shape even
if the accretion takes place on two poles. However, we point out
the very high significance level of the 24 min peak in September
7 night and the evident even-odd asymmetry in the folded light
curve in Fig.~\ref{fig6}.

In conclusion, in spite that the evidence for this longer period
is present only in one observation run and in spite that the run
cover only less than 5 periods, we tend to identify the 24 min
period as the true spin period, but further observations (in
particular, time resolved spectroscopy) are needed in order to
confirm this outcome.

In the time series from October 1 observation, the modulation at
12 min is very low and the corresponding peak in the periodogram
is well below the 90\% confidence level. In the same night, the
mean flux is 20\% lower than the flux averaged on the June
observations and in the middle of the run a dip is present: a
possible explanation of this behavior could be a partial eclipse
of the accreting
disc. \\
Regarding the long term variability, we are not able to ascribe
it to an orbital modulation, because the low frequency peaks in
the DFT correspond to long periods, poorly sampled by the
observations. The presence of power at low frequency could be
produced by instabilities in the accretion disc or curtain.

\section{Conclusions}
We have presented time resolved optical photometry of the stars
included in the field of the X-ray pulsating source \object{1WGA
J1958.2+3232}. A stable, high amplitude, periodic modulation was
found in the light curves of the star previously proposed by
Israel et al. (\cite{Isr99}) as the optical counterpart of
\object{1WGA J1958.2+3232}. The 12 min period, compatible with
the X-ray one, is likely associated with the first harmonic of
the spin of a white dwarf in an Intermediate Polar Cataclysmic
Variable. If this outcome will be confirmed, \object{1WGA
J1958.2+3232} would be, among the Intermediate Polars, the slowest
rotator which exhibit double peaked spin pulse shape.

\begin{acknowledgements}
The authors are grateful to Domitilla de Martino, Gian Luca
Israel, Sandro Mereghetti, Pablo Reig and Luigi Stella for helpful
conversations and to Lucio Chiappetti and Constantinos Paizis for
a critical reading of the manuscript. Financial support from the
Catania Astrophysical Observatory and the \textit{Regione
Sicilia}, is gratefully acknowledged. This work was partially
supported by the European Commission under contract ERB
FMRXCT98-0195 and by \textit{Agenzia Spaziale Italiana} (ASI).
\end{acknowledgements}

\end{document}